\documentclass[pre,showpacs,twocolumn,floatfix,bibnotes]{revtex4}

\usepackage{here}
\usepackage[dvips]{graphicx}

\newcommand\pictc[5]{\begin{figure}
                       \centerline{
                       \includegraphics[width=#1\columnwidth]{#3}}
                   \protect\caption{\protect\label{fig:#4} #5}
                    \end{figure}            }
\newcommand\pict[4][1.]{\pictc{#1}{!tb}{#2}{#3}{#4}}
\newcommand\rpict[1]{\ref{fig:#1}}
\newcommand\leqt[1]{\protect\label{eq:#1}}
\newcommand\reqtn[1]{\ref{eq:#1}}
\newcommand\reqt[1]{(\reqtn{#1})}

\newcounter{Fig}

\begin{document}

\begin{sloppy}

\title{Subwavelength imaging with opaque left-handed nonlinear lens}

\author{Alexander A. Zharov$^{1,2}$, Nina A. Zharova$^{1,3}$, Ilya V. Shadrivov$^1$ and Yuri S. Kivshar$^1$}

\affiliation{$^1$ Nonlinear Physics Group, Research School of
Physical Sciences and Engineering, Australian National
University, Canberra ACT 0200, Australia \\
$^2$ Institute for Physics of Microstructures, Russian Academy of
Sciences, Nizhny Novgorod 603950, Russia\\
$^3$ Institute of Applied Physics, Russian Academy of Sciences,
Nizhny Novgorod 603600, Russia}

\begin{abstract}
We introduce the concept of subwavelength imaging with an opaque
{\em nonlinear left-handed lens} by generating the second-harmonic
field. We consider a slab of composite left-handed metamaterial
with quadratic nonlinear response and show that such a flat lens
can form, under certain conditions, an image of the
second-harmonic field of the source being opaque at the
fundamental frequency.
\end{abstract}


\maketitle

One of the most unique properties of the recently demonstrated
left-handed metamaterials, i.e. materials with simultaneously
negative real parts of dielectric permittivity and magnetic
permeability~\cite{Veselago:1967-517:UFN}, is their ability to
focus electromagnetic waves by a flat slab of the material, the
property which makes these materials quite different from the
conventional optical lenses with the positive refractive index
needed to have curved surfaces to form an image. Recently,
Pendry~\cite{Pendry:2000-3966:PRL} argued that a slab of a
lossless left-handed material with $\epsilon = \mu =-1$ should
behave like {\em a perfect lens} enabling to obtain an ideal
image of a point source through the amplification of the
evanescent components of the field.

While recent experimental demonstrations confirmed the main
features of negative refraction of the left-handed
materials~\cite{exp1,exp2}, the question of near-perfect imaging
by flat lens and near-field focusing still remain highly
controversial~\cite{contr1}, and it severely constrained because
of large dissipation and dispersion of metamaterials.
Nevertheless, numerical studies indicate~\cite{cummer} that
nearly-perfect imaging should be expected even under realistic
conditions when both dispersion and losses of the left-handed
composites are taken into account. More importantly, some of the
properties of the left-handed materials, such as {\em negative
refraction}, have been predicted~\cite{theory} and recently
demonstrated experimentally~\cite{PC_exp} in {\em photonic
crystals}, which are inhomogeneous periodic structures with a
lattice constant comparable to the wavelength. This negative
refraction allows considerable control over light propagation, and
it opens up doors for new applications of microstructured
materials and the extension of the basic concepts from microwaves
to optical frequencies.

Until now, all properties of flat lenses and subwavelength imaging
have been studied only for linear waves. However, it has been
already noticed that the left-handed metamaterials may possess
quite complicated nonlinear magnetic
response~\cite{Zharov:2003-37401:PRL}, or properties of such
materials can be altered by inserting diodes in the resonators of the composite structure thus making the response of the entire material nonlinear~\cite{Lapine:2003-65601:PRE}. In
this Letter, we make one step forward into this direction and
study the effects of the second-harmonic generation and
subwavelength imaging by a nonlinear left-handed lens. In
particular, we analyze the imaging properties of a slab of
metamaterial with quadratic nonlinear response and demonstrate,
both analytically and numerically, that such a slab can form an
image of the second-harmonic field of the source being opaque at
the fundamental frequency. This can happen under certain
conditions, which include Pendry's conditions of a perfect lens
satisfied for the second-harmonic field, $\epsilon(2\omega)=\mu(2
\omega)=-1$, leading to the conditions for $\epsilon (\omega)$
and $\mu(\omega)$ being of the opposite signs at
the fundamental frequency $\omega$. More importantly, for the case
of two sources we show that the resolution of such a nonlinear
left-handed flat lens can be  made indeed better than the
radiation wavelength.

We consider a lens in the form of a layer of left-handed
metamaterial with the thickness $D$, as shown schematically in
Fig.~\rpict{schem}. We assume that the metamaterial is a
three-dimensional composite structure made of wires and split-ring
resonators (SRRs) in the form of a cubic lattice. When the lattice
period $d$ is much smaller than the radiation wavelength
$\lambda$, i.e. $d \ll \lambda$, this composite structure can be
described within the effective-medium approximation, and it can be
characterized by dielectric permittivity and magnetic permeability
which, for the specific structure and in the linear regime can be
derived consistently and can be written in the form
\begin{equation}\leqt{epsilon_eff}
\epsilon(\omega) = 1 - \frac{\omega_p^2}{\omega(\omega - i\gamma_e)},
\end{equation}
\begin{equation}\leqt{mu_eff}
\mu(\omega) = 1 + \frac{F
\omega^2}{\omega_0^2-\omega^2+i\gamma_m\omega},
\end{equation}
where $\omega_p = \sqrt{2\pi c^2 / d^2 \ln{(d/r)}}$ is the
effective plasma frequency, $\omega_0 =
\bar{\omega_0}\sqrt{(1-F)}$, $\bar{\omega_0}$ is the
eigenfrequency of an isolated SRR, $F$ is the volume density of
SRRs, $\gamma_e$ and $\gamma_m$ are the damping coefficients,
$\omega$ is the frequency of the external electromagnetic field,
$r$ is the wire radius, and $c$ is the speed of light. In the
frequency range where the real parts of $\epsilon$ and $\mu$ are
both negative and for $\gamma_e, \gamma_m \ll \omega$, such a
composite structure demonstrates left-handed transmission, whereas
for $\omega < \omega_0$, it is opaque because the signs of
$\epsilon$ and $\mu$ are opposite.

\pict{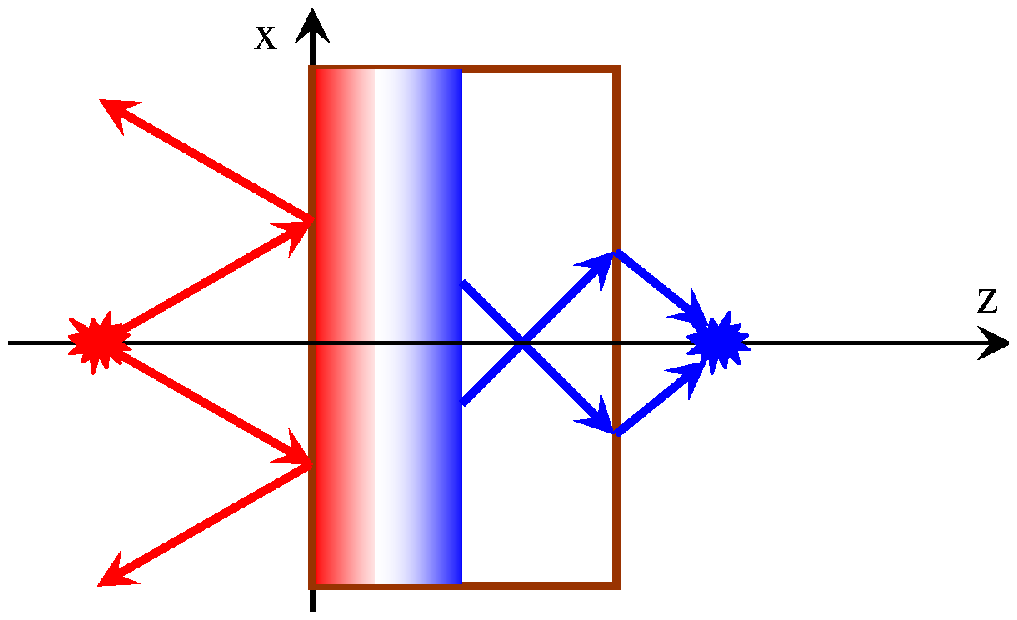}{schem}{Schematic of the problem. Electromagnetic
waves emitted by a source (located at $z=-z_s$) are reflected
from an opaque slab of a left-handed material. Inside the slab,
the exponentially decaying field at the fundamental frequency
$\omega$ generates the second-harmonic field at $2 \omega$, which
penetrates through the slab creating an image at the second
harmonics.}

In order to achieve a nonlinear response in such a structure, we
assume that each SRR includes a nonlinear element, e.g. a diode
inserted in the SRR~\cite{Lapine:2003-65601:PRE}. If the diode has
an asymmetric current-voltage characteristics, the unit cell does
not possess a center of symmetry with respect to the field
direction, and the resulting nonlinear response should include, in
particular, the second harmonic of the source frequency.

Our idea is to satisfy the perfect-lens conditions for the
second-harmonic waves and, therefore, we take $\mu(2\omega) = \epsilon(2\omega) = -1$. From Eqs.~(\ref{eq:epsilon_eff}),
(\ref{eq:mu_eff}) we can find that these conditions can be
satisfied provided
\begin{equation}\leqt{ff_param}
\epsilon(\omega)=-7, \;\;\;\;\;\; \mu(\omega) =(3-F)/(3-2F).
\end{equation}
For this choice of the material parameters, the slab is opaque at
the fundamental frequency $\omega$, and the waves do not penetrate
into it. However, the nonlinear quadratic response of the material
can allow the process of the second-harmonic generation. Since the
material is transparent at the frequency $2 \omega$, we expect
that the second-harmonic field can propagate through the slab
creating an image of the source behind the flat lens.

Using the so-called undepleted pump approximation, we can obtain
the equation for the TM-polarized second-harmonic field
$H_y^{(2\omega)}(x,z)$ inside the slab, which has the form
well-known in the theory of the second-harmonic generation (see
e.g. Ref.~\cite{Vinogradova:1990:TheoryWaves})
\begin{equation}\leqt{H_nl}
\Delta H_y^{(2\omega)} + K^2(2\omega) H_y^{(2\omega)} =
        -\frac{16 \pi \omega^2 \epsilon(2\omega)}{c^2}M_{NL}^{(2\omega)},
\end{equation}
where $\Delta$ is the Laplacian acting in the space $(x,z)$,
$K^2(2\omega) = 4 k_0^2 \epsilon(2\omega)\mu(2\omega)$, and
$M_{NL}^{(2\omega)}$ is the nonlinear magnetization of the unit
volume of the metamaterial at the frequency $2\omega$, which
appears due to the nonlinear magnetic momentum of SRR,
\begin{equation}
\leqt{M_nl} M_{NL}^{(2\omega)} = - \frac{3}{4} \chi(\omega)
\left[H_y^{(\omega)}(x,z)\right]^2,
\end{equation}
where
\[
\chi(\omega) = \frac{(\pi a^2)^3 \omega_0^4}{c^3 d^3 U_c R_d
\omega^2} \left[ \left(\frac{\omega_0^2}{\omega^2}-1\right)^2 -
        i\frac{\gamma_m}{\omega} \left(\frac{\omega_0^2}{\omega^2}+2
        \right)
    \right]^{-1},
\]
$H_y^{(\omega)}(x,z)$ is the spatial distribution of the magnetic
field at the fundamental frequency in the slab, $a$ is the radius
of the resonator rings, $R_d$ is the differential resistance of
the diode at zero voltage, $k_0=\omega/c$, and $U_c$ is the diode
parameter defined from the current-voltage characteristics
of the diode which we take in the form $I=I_0 (e^{U/U_c} - 1)$.
The right-hand side of Eq.~\reqt{H_nl} vanishes outside the metamaterial slab.

Applying the Fourier transform along the $x$ direction,  we obtain
the equation for the function
$\bar{H}_y^{(2\omega)}=\bar{H}_y^{(2\omega)}(k_x,z)$ in terms of
$\bar{H}_y^{(\omega)}=\bar{H}_y^{(\omega)}(k_x,z)$ and $G =
G(k_x,z)$ which are the Fourier transforms of $H^{(\omega)}(x,z)$,
$H^{(2\omega)}(x,z)$ and  $[H^{(\omega)}(x,z)]^2$, respectively:
\begin{equation}\leqt{H_nl_kx}
\frac{d^2 \bar{H}_y^{(2\omega)}}{dz^2} +
        \left[K^2(2\omega)-4k_x^2\right] \bar{H}_y^{(2\omega)} =
        \eta G,
\end{equation}
where $ \eta = 12 \pi k_0^2 \epsilon(2\omega) \chi(\omega)$.
Using the convolution theorem, we express the function
$G$ through the spectrum of the magnetic
field at the fundamental frequency $\bar{H}_y^{(\omega)}$ in the
form
\begin{equation}\leqt{convolution}
 G = \int_{-\infty}^{\infty}
        \bar{H}^{(\omega)}(k_x^{\prime},z) \bar{H}^{(\omega)}(k_x-k_x^{\prime}, z) \, dk_x^{\prime}.
\end{equation}
Within the framework of the undepleted pump approximation, the
function $\bar{H}^{(\omega)}(k_x,z)$ can be found as a solution of
the linear problem describing the electromagnetic field at the
fundamental frequency transmitted into the left-handed slab,
\begin{equation}\leqt{spectrum}
\bar{H}^{(\omega)} = \frac{2 \kappa_1 e^{- i k_0 \kappa_1 z_s}}{
D_s(\omega,k_x)} \left[ Z_{1}e^{k_0 \kappa_2 z} -
    Z_{2} e^{k_0\kappa_2 (2D -z)} \right] S(\gamma),
\end{equation}
where $S(\gamma)$ is the spectral function of the source at the
fundamental frequency located at the distance $z_s$ from the left-handed slab, $\kappa_1 = \sqrt{1-\gamma^2}$, $\kappa_2
= \sqrt{\gamma^2 - \epsilon(\omega)\mu(\omega)}$, $\gamma =
k_x/k_0$, $Z_{1,2} = \kappa_1 \pm i\kappa_2/ \epsilon(\omega)$,
and $D(\omega,k_x) = Z_1^2-Z_2^2\exp{(2 k_0 \kappa_2 D)}$. The
spectral function $S(\gamma)$ includes both fast propagating
($\gamma \le 1$) and slow  evanescent ($\gamma > 1$) spatial
harmonics. Possible distortions of the second-harmonic image can
be caused by the pre-exponential factor in Eq.~\reqt{spectrum},
and the main effect is due to the pole singularity defined by the equation
\begin{equation}\leqt{pole}
 D(\omega,k_x) = 0,
\end{equation}
that characterizes the resonant excitation of {\em surface
polaritons} which are known to limit the resolution of a left-handed lens. In the case of a thick slab, i.e. $k_0 \kappa_2 D
\gg 1$, the resonant wavenumber of the surface waves can be
found in the form
\begin{equation}\leqt{surface_wave}
\gamma^2_{\rm sp} =
\frac{\epsilon(\omega)[\epsilon(\omega)-\mu(\omega)]}{\epsilon^2(\omega)-1}.
\end{equation}
Substituting the explicit expressions for $\epsilon(\omega)$ and
$\mu(\omega)$ from Eq.~\reqt{ff_param} into Eq.~\reqt{surface_wave}, we obtain a simple
estimate for an expected resolution limit of the nonlinear
left-handed lens in terms of the critical (limiting) wavenumber,
\begin{equation}\leqt{res_limit}
\gamma^2_{\rm lim} \approx 1 + \frac{(3-F)}{7(3-2F)}.
\end{equation}
However, the existence of this critical wavenumber does not
necessarily limits the lens resolution and, in reality, the effect
of surface waves on the imaging properties of the nonlinear
lens depends on the efficiency of their excitation by each
particular source.

Analytical solution of the problem for the spatial spectrum of the
second-harmonic field transmitted through the left-handed slab can
be obtained for narrow enough spectrum of the source, i.e., when
the width of the source spectrum at the fundamental frequency does
not exceed the value $\gamma_c$, where
\begin{equation}
\gamma_c^2 \approx |\epsilon(\omega) \mu(\omega)|.
\end{equation}
Then, we can use the impedance boundary conditions for the
fundamental field at the interface between vacuum and the
metamaterial slab at $z = 0$. Subsequent numerical results
indicate that this approximation remains valid provided $\gamma_c
\gg \gamma_{\rm lim}$.

To solve the problem analytically, we assume that the wave at the fundamental frequency $\omega$ penetrates inside the slab on a distance (the skin layer) much smaller than the slab thickness $D$, i.e. $D
\delta \gg 1$, where $\delta = k_0\sqrt{-\epsilon(\omega)\mu(\omega)}$. Taking into account the actual values of $\epsilon(\omega)$ and $\mu(\omega)$ \reqt{ff_param}, one can see that the fundamental frequency penetration depth, $\sim (2\pi\delta)^{-1}$, does not exceed $\lambda/17$. Then, Eq.~\reqt{H_nl_kx} can be re-written in the form
\[
\frac{d^2 \bar{H}_y^{(2\omega)}}{dz^2} +
        \left[K^2(2\omega) - 4k_0^2\gamma^2\right]
        \bar{H}_y^{(2\omega)} =
        \eta e^{-2\delta z} A_0(\gamma),
\]
where $A_0(\gamma) = \int_{-\infty}^{\infty}
\xi(\gamma^{\prime})\xi(\gamma - \gamma^{\prime}) d
\gamma^{\prime}$, and $\xi(\gamma) = e^{\delta z} \bar{H}^{(\omega)}(\gamma,z)$ does not depend on $z$. As a result, the general solution for the second harmonic inside the slab can be presented in the form
\begin{equation}\leqt{H_2w_kx_3}
\bar{H}_y^{(2\omega)}(\gamma,z) = C_1 e^{2 k_0 \kappa_2 z} +
    C_2 e^{-2 k_0 \kappa_2 z} +
    C_3 e^{-2 \delta z},
\end{equation}
where $C_{1,2}$ are two constants which should be determined from the
boundary conditions, and
\begin{equation}\leqt{C_3}
C_3 = \left(\frac{1}{4k_0^2}\right) \frac{\eta
A_0(\gamma)}{[\epsilon(2\omega)\mu(2 \omega) -
\epsilon(\omega)\mu(\omega)]}.
\end{equation}

\pict{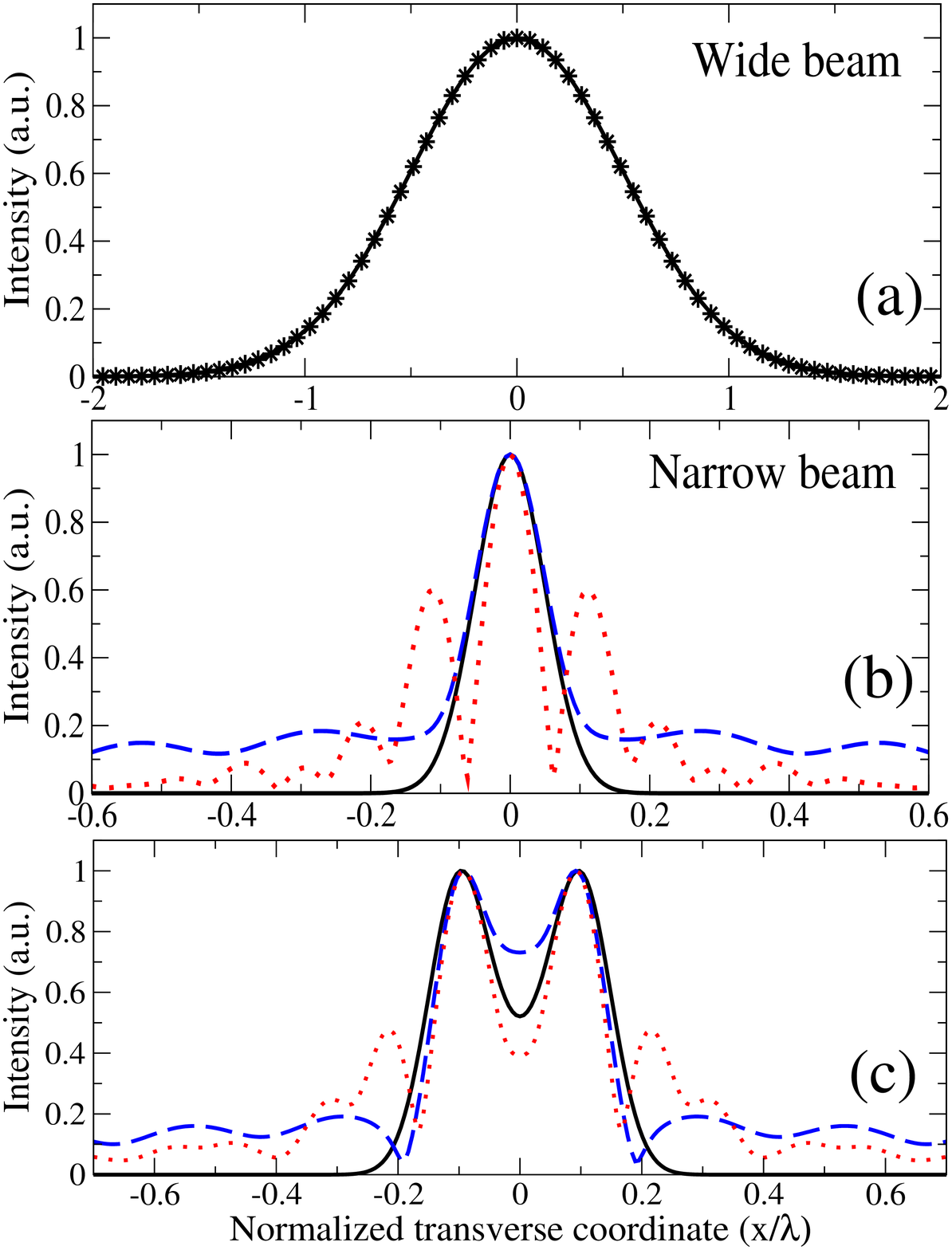}{imag}{Numerical results for imaging by a nonlinear
left-handed lens. Shown are the intensities of the fundamental
field at the source location (solid) and the second-harmonic field
at the image plane for different cases (stars, dashed and dotted
lines). (a) Wide beam (the width is $\lambda$) generated by a
single source, $D=\lambda$, $z_s =\lambda/2$. (b) Narrow beam (the
width is $\lambda/10$) generated by a single source; dashed--the
image for $D=\lambda/10$, $z_s = \lambda/20$; dotted -- the image
for $D=0.3\;\lambda$, $z_s = 0.15 \; \lambda$. (c) Imaging by two
sources separated by the distance $\lambda/5$; dashed -- the image
for $D=\lambda/10$ and $z_s = 0.03 \; \lambda$;  dotted -- the
image for $D=0.3\; \lambda$ and $z_s = \lambda / 5$.}

We should satisfy the continuity of the tangential components of
the magnetic $H_y^{(2\omega)}$ and electric $E_x^{(2\omega)}$
fields at the interfaces between air and the metamaterial slab, i.e.
at $z=0$ and $z=D$.  As a result, we obtain the second harmonic of
the magnetic field behind the slab (for $z>D$) where the image is
expected to form,
\begin{equation}\leqt{transm_spectr}
\bar{H}_y^{(2\omega)}(\gamma, z) = -\frac{1}{2} \left\{  1 - i
\frac{\kappa_2 \epsilon(\omega)}{\kappa_1 \epsilon(2\omega)}
\right\} C_3 e^{ 2 k_0\kappa_1 (2 D - z)}.
\end{equation}
For wide beams with narrow spectra, Eq.~\reqt{transm_spectr} can
be rewritten in the form
\begin{eqnarray}\leqt{transm_spectr_2}
\bar{H}_y^{(2\omega)}(\gamma, z) =
    -\frac{\eta}{2 k_0^2[\epsilon(2\omega)\mu(2\omega) -
        \epsilon(\omega)\mu(\omega)]} \times \nonumber \\
\left[
    1 - \frac{\sqrt{-\epsilon(\omega)\mu(\omega)}}{\epsilon(2\omega)}
\right]
\left[
    1 - i\frac{\sqrt{-\epsilon(\omega)\mu(\omega)}}{\epsilon(\omega)}
\right]^{-1} \times \nonumber\\
\exp{\left\{ 2k_0\kappa_1 (2D-z-z_s)\right\}}
\int_{-\infty}^{\infty} S(\gamma^{\prime})S(\gamma - \gamma^{\prime}) \,
d\gamma^{\prime}.
\end{eqnarray}

Thus, the squared field at the fundamental frequency acts as an
effective source of the second-harmonic field and, as a result,
the image of the squared field is reproduced by the nonlinear
left-handed lens. This image appears at the point $z_{\rm
im}=D-z_s$, and this result coincides with the corresponding
result for the linear lens discussed
earlier~\cite{Pendry:2000-3966:PRL}.

When the size of the source is comparable or less than the
wavelength $\lambda$ of the fundamental-frequency wave, the
problem cannot be treated analytically, and Eq.~\reqt{H_nl} has
been solved numerically. In Figs.~\rpict{imag}(a-c), we present
our numerical results for the intensity distribution of the
incident beam at the source point and the field distribution of
the second-harmonic beam at the image location, normalized to the
field maxima. The actual amplitude of the electromagnetic field at
the image location is lower than the amplitude of the source
because of a finite efficiency of the process of the
second-harmonic generation. For the objects with the spatial scale
larger or equal to the radiation wavelength, the second-harmonic
field profile coincides with the intensity of the fundamental
field generated by the source, as shown in Fig.~\rpict{imag}(a).

\pict{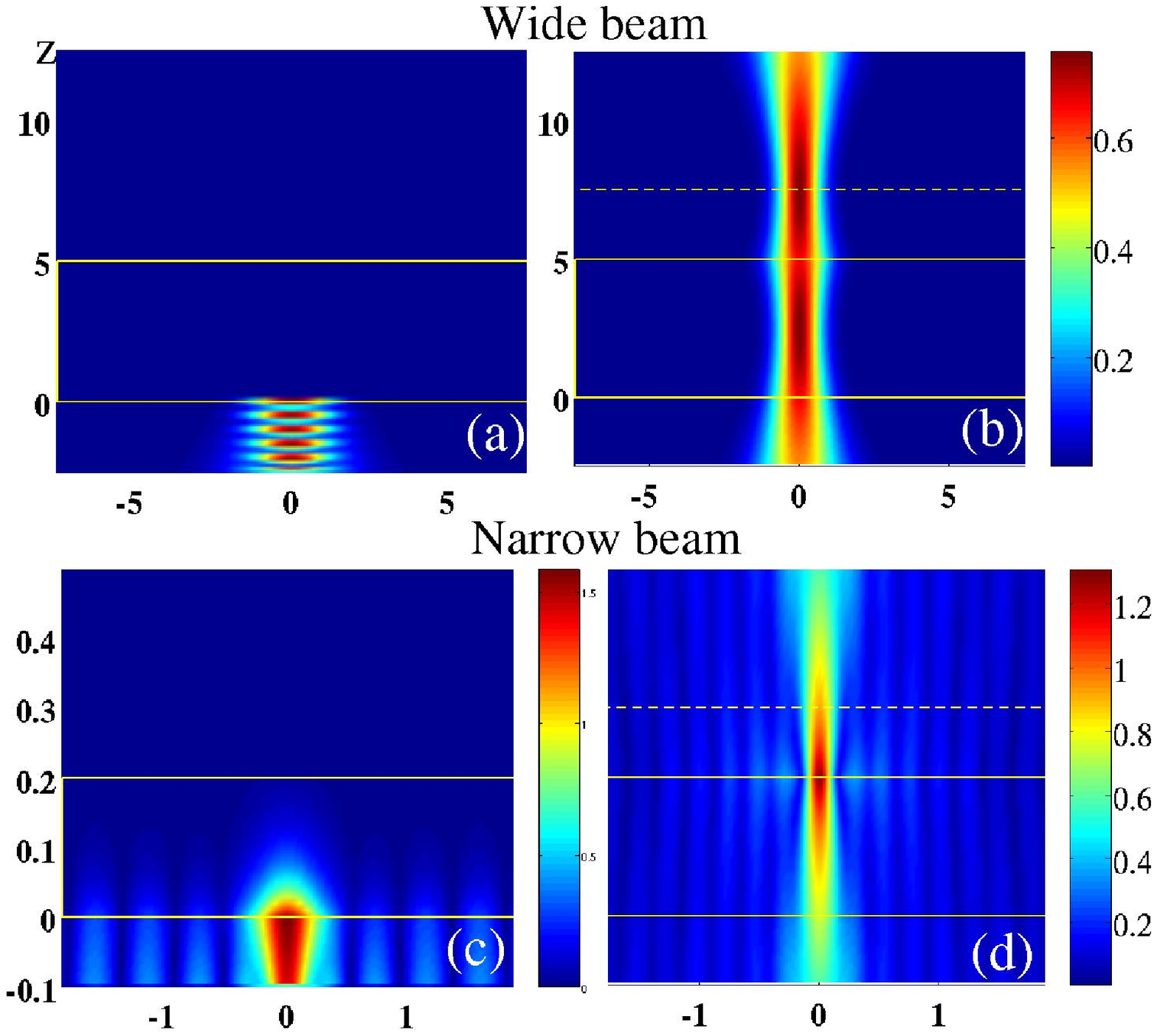}{combined}{Intensity of the fundamental [(a,c)] and
second-harmonic [(b,d)] beams (in units of the wavelength) for the
problem of the second-harmonic generation and imaging by a
nonlinear left-handed lens. (a,b) Wide beam ($D = 5 \lambda$, $z_s
= 2.5 \lambda$, and $a_0 = \lambda$) and (c,d) narrow beam ($D =
 \lambda /5$, $z_s = \lambda/10$, and $a_0 = \lambda/4$). Solid
lines mark the flat surfaces of the nonlinear left-handed lens.
Dashed lines show the predicted locations of the second-harmonic
image.}

However, when the source contains the spatial scales less than the
radiation wavelength, the imaging properties of the nonlinear lens
depend strongly on the slab thickness $D$. As an example, in
Fig.~\rpict{imag}(b) we show the results for the transmission of
an incident Gaussian beam of the width $\lambda/10$ which
reproduces almost exactly the source profile at the image plane in
the case of a thin lens (dashed line) but generates a strongly
distorted image when the slab thickness becomes larger than a half
of the wavelength $\lambda$. Distortions appear as periodic
variation of the second-harmonic field being caused by excitation
of surface waves. Intensity distribution of the magnetic field in the
fundamental and second-harmonic fields are shown in
Figs.~\rpict{combined}(a-d) for (a,b) large and (c,d) small
(compared to the radiation wavelength) size of the source,
respectively.

Figure~\rpict{imag}(c) shows the numerical results for imaging of two
sources that generate the Gaussian beams with the maxima
separated  by the distance $\lambda/5$ in the transverse
direction. Again, the image reproduces very well the source for a
thinner lens, and therefore a thin nonlinear lens does provide a
subwavelength resolution of the second-harmonic field. In contrast
to the linear flat lens, the resolution of the nonlinear lens
depends on the distance $z_s$ between the source and the lens, and
the optimal distance can be determined separately for each
particular case.

In conclusion, we have studied the second-harmonic generation and
subwavelength imaging by a nonlinear left-handed lens, i.e. a slab
of metamaterial with quadratic nonlinear response. We have
demonstrated, both analytically and numerically, that such a slab
can form an image of the second-harmonic field of the source being
opaque at the fundamental frequency, with the resolution that can
be made indeed better than the radiation wavelength.

\end{sloppy}
\end{document}